# REVIEW OF INTELLIGENT TUTORING SYSTEMS USING BAYESIAN APPROACH

**Dr.R. Santhi**
Dean – Student Affairs, Head-ERP, Alpha College of Engineering, Chennai-116
rshanthi.teacher@gmail.com

**B.Priya**
Asst Prof Gr – II, MCA Dept, Sri Sai Ram Engineering College, Chennai.
priya.mca@sairam.edu.in

**J.M.Nandhini**
Asst Prof Gr – I, MCA Dept, Sri Sai Ram Engineering College, Chennai.
nandhini.mca@sairam.edu.in

## Abstract

With advancement in computer science research on artificial intelligence and in cognitive psychology research on human learning and performance, the next generation of computer-based tutoring systems moved beyond the simple presentation of pages of text or graphics. These new intelligent tutoring systems (ITSs) called cognitive tutors; incorporated model-tracing technology which is a cognitive model of student problem solving that captures students' multiple strategies and common misconceptions. Such Intelligent tutoring systems or Knowledge Based Tutoring Systems can guide learners to progress in the learning process at their best. This paper deals with the review of various Intelligent tutoring systems using Bayesian Networks and how Bayesian Networks can be used for efficient decision making.

## 1. Introduction

An Intelligent Tutoring Systems is educational software containing an Artificial Intelligence component. The software tracks the student's work, tailoring feedback and hints along the way. By collecting information on a particular student's performance, the software can make inferences about strengths and weaknesses, and can suggest additional work. The basic underlying idea of ITSs is to realize that each student is unique. These systems can be used in the traditional educational setting or in distant learning courses, either operating on stand-alone computer or as applications that deliver knowledge through the internet.

ITSs have been shown to be effective in increasing student's performance and motivation levels compared with traditional instructional methods. One of the key element that distinguishes ITSs from more traditional CAI systems is ITS's capability to dynamically maintain a model of a student's knowledge during the study.

ITS must be able to achieve three main tasks:

1. Accurately diagnose a student's knowledge level using principles rather than programmed responses.
2. Decide what to do next and adapt instruction accordingly.
3. Provide feedback.

This kind of diagnosis and adaptation, which is usually accomplished using Artificial Intelligences, is what distinguishes ITS from CAI. Bloom demonstrates that individual one-on-one tutoring is the most effective mode of teaching and learning [1].

## 2. Emergence of ITS

In 1982, Sleeman and Brown reviewed the state of the art in computer aided instruction and first coined the term Intelligent Tutoring Systems (ITS) to describe these evolving systems and distinguish them from the previous CAI systems. With new AI techniques coming up it seemed that the computers were almost capable of "thinking" like humans. This motivated ITS research further. Application of AI in ITS made it possible to achieve the goals more easily. Other reasons which motivated ITS research were[2]:

- Modular and fine grained the curriculum.
- Customized for different student populations.
- Individual presentation and assessment of the content.
- Collection of data which instructors could use to tutor and remediate students

ITS research has successfully delivered techniques and systems that provide adaptive support for student problem solving or question-answering activities in a variety of domains (e.g., programming, physics, algebra, geometry, SQL and introductory computer science)

## 3. Architecture of an ITS

There are four emerging subsystem for an ITS, namely:

1. Knowledge Base Model
2. Student Model
3. Teaching Model
4. Control Engine

Two major issues related to an ITS are "what to teach" and "how to teach". The Knowledge Base Model deals with the "what to teach" part,

whereas the Student and the Teaching model are concerned with "how to teach" part.

The Knowledge Base model is also important as it is the representation of the domain knowledge. Good design principles adapted in designing the Knowledge Base model would help the system in selecting the proper methods for teaching. This would also help the system in the search for alternative teaching plans when a particular method fails to produce success. The most important part of an ITS is the teaching model. This module is the centre of the whole system. It communicates with the other modules and does the entire decision making.

In our proposed system, a Control Engine agent visualizes the Knowledge Base Model and the Student Model to analyze the various learning parameters of the learner such as the allotted time for studying a specific content. Based on the strategy of the Teaching Model, the learner at any time is able to choose the learning contents himself which leads to the updation of the Student Model as illustrated in Figure 1.

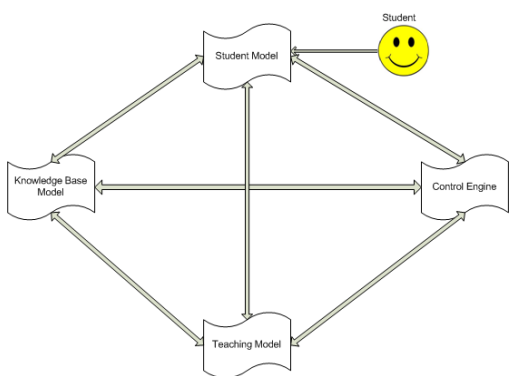

Figure 1: Architecture of ITS

## 4. Methods of handling Uncertainty

One of the biggest challenges in designing ITSs is the effective assessment and representation of the student's knowledge state and specific needs in the problem domain based on uncertainty information.

The task of dealing with the uncertainty management for the student model is thus challenging. Various approaches in Artificial Intelligence have been proposed for uncertainty reasoning , including

1. Rule-based systems
2. Fuzzy logic
3. Dempster Shafer theory of Evidence
4. Neural networks and
5. Bayesian networks

Bayesian networks are a powerful approach for uncertainty management in Artificial Intelligence .

## 5. Bayesian Networks

The recent trend in ITS involves the creation of systems that can make decisions based on uncertain or incomplete information. One formal framework for uncertainty management is Bayesian Networks which use probability theory as a formal framework for uncertainty management in Artificial Intelligence.[5]

A Bayesian Network (BN) is a graphical description of a probability distribution that permits efficient probability propagation combined with a rigorous formalism. A BN for a given domain represents the joint probability distribution, p(x), over the set of random variables, X, of the domain, as a set of local distributions combined with a set of conditional independence assertions.

Researchers have applied Bayesian networks to many tasks [6], including

1. Student Modeling
2. E-Commerce and
3. Multi Agents

In student Modelling, there are two tasks involved in helping a student navigate in a personalized learning environment.

1. The structure of the problem domain must be modeled.
2. Student knowledge regarding each concept in the problem domain should be tracked.

Bayesian Networks can help us meet both these objectives.

A Bayesian network (BN) consists of directed acyclic graph (DAG) and a corresponding set of conditional probability distributions (CPDs). Based on the probabilistic conditional independencies encoded in the DAG, the product of the CPDs is a joint probability distribution (jpd). In other words, Bayesian networks serve as both a semantic modeling tool and an economical representation of a jpd. There are many inference algorithms in BNs for computing probabilities of variables given other variables to take on certain values [7].

A Bayesian Network combines the contents structure with the user profile and learning style to suggest pedagogical directions.

Figure 2 plots a sample Bayesian Network where variables **B** and **D** are independent, variable **C** is only dependent of **D**, and variable **A** is conditionally independent of the variable **D** given **B** and **C** variables.

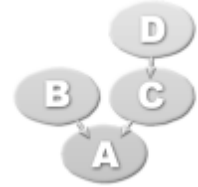

Figure 2: A Bayesian Network

In the context of ITSs, Bayesian Networks have been applied to user modeling (VanLehn et al. 1998) in a diagnostic perspective: given a student action (symptom) the network provides the most likely state of knowledge (diagnosis)[8].

## 6. Literature Review

Andes (Conati et al, 2002; Gertner & VanLehn, 2000) is an ITS which was developed to teach physics for the students in Naval Academy. Bayesian networks were primarily used in Andes for decision making.

The major foci of the system are

(1) Select the most suitable strategy for the student

(2) Predict Student's actions

(3) Perform a long term assessment of the student's domain knowledge.

Andes is a domain dependent ITS. Each problem in the system was broken into some steps and Bayesian network was formed using those steps as nodes. So, the problems were represented in the system as Bayesian networks. The Bayesian network would predict the most probable path for the student during a course. Each student could have different approaches to a problem, the network would be adjusted accordingly (the probabilities would change) and finally for a new problem it would predict the best strategy for the student[4].

ViSMod is another ITS which used Bayesian network (Zapata-Rivera et al, 2004). In the system the Bayesian network was divided into three levels. At the top most level the concepts (to be taught) were represented in a hierarchical manner. After that in the second level student's performance and behavior were described. Finally the third level nodes represented some analysis on the student's performance. Only the first level is domain dependent, whereas other two levels would remain same over different domains. Again student can observe only the top two levels of the Bayesian net. The third level is only visible to the teachers. During a course the probabilities in the second and third level of the Bayesian network changed according to the student's performance.

BITS, a web based Intelligent tutoring system for Computer Programming, uses Bayesian Networks for making the decisions. Using the Bayesian Network, the prerequisite relationships among the concepts are represented directly and clearly. In BITS, there are two methods of obtaining the evidence required to update the Bayesian Network:

A student's direct reply to a BITS query if this student knows a particular concept.

A sample quiz result for the corresponding concept to determine whether the student has understood a particular concept or not[3].

### 7. Case Study on Data Structures- Graphs

Graph is denoted by {V,E} where V is the set of vertices and E is the set of edges connecting the vertices. The topics related to Graph can be represented by a Knowledge Dependency Graph in the Knowledge Base Model as:

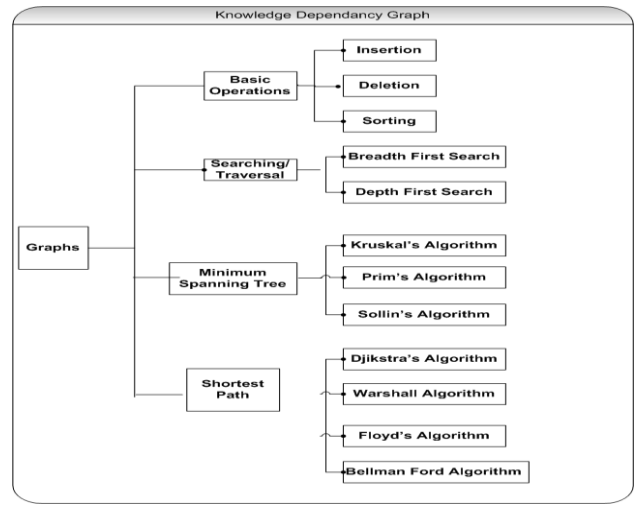

Figure 3 Knowledge Dependency Graph

As seen in Figure 3, the prerequisite for studying the various concepts can be stored in the Knowledge Base model to facilitate easy understanding to the learners.